\documentclass[aps, prl, twocolumn, amsfonts, amsmath, amssymb, floatfix]{revtex4}

\usepackage[T1]{fontenc}
\usepackage[latin9]{inputenc}
\usepackage{color}
\usepackage{graphicx}
\usepackage{url}
\usepackage[normalem]{ulem}
\usepackage{amsthm}
\usepackage{amsmath}

\def\be{\begin{equation}}
\def\ee{\end{equation}}
\def\bea{\begin{eqnarray}}
\def\eea{\end{eqnarray}}

\begin{document}

\title{Anderson localization and Mott insulator phase in the time domain}

\author{Krzysztof Sacha} 
\affiliation{
Instytut Fizyki imienia Mariana Smoluchowskiego, 
Uniwersytet Jagiello\'nski, ul. Prof. S. \L{}ojasiewicza 11 PL-30-348 Krak\'ow, Poland}
\affiliation{Mark Kac Complex Systems Research Center, Uniwersytet Jagiello\'nski, ul. Prof. S. \L{}ojasiewicza 11 PL-30-348 Krak\'ow, Poland}

\begin{abstract}
Particles in space periodic potentials constitute standard models for investigation of crystalline phenomena in solid state physics. Time periodicity of periodically driven systems is a close analogue of space periodicity of solid state crystals. There is an intriguing question if solid state phenomena can be observed in the time domain. Here we show that wave-packets localized on resonant classical trajectories of periodically driven systems are ideal elements to realize Anderson localization or Mott insulator phase in the time domain. Uniform superpositions of the wave-packets form stationary states of a periodically driven particle. However, an additional perturbation that fluctuates in time results in disorder in time and Anderson localization effects emerge. Switching to many-particle systems we observe that depending on how strong particle interactions are, stationary states can be Bose-Einstein condensates or single Fock states where definite numbers of particles occupy the periodically evolving wave-packets. Our study shows that non-trivial crystal-like phenomena can be observed in the time domain. 
\end{abstract}

\maketitle


Nearly sixty years ago Anderson discovered that transport of non-interacting particles in the presence of disorder can stop totally due to interference of different scattering paths \cite{Anderson1958}. The suppression of the transport is accompanied by localization of eigenstates in the configuration space \cite{tiggelen99,Lagendijk2009}. Quantum description of classically chaotic systems reveals yet another version of Anderson localization (AL) where localization of a particle takes place in the momentum space \cite{haake2010}. Suppression of classical chaotic diffusion in the momentum space was identified with AL after the quantum kicked rotor system was mapped to the quasi-random one-dimensional (1D) Anderson model \cite{fishman82}. Here we show that the presence of time disorder in periodically driven systems can induce AL in the time evolution.

Controlling interactions between bosonic particles in the presence of a space periodic potential allows for investigation of quantum phase transitions \cite{sachdev}. In the limit of weak interactions, particles in the ground state 
reveal long-range phase coherence. For strong repulsive contact interactions, the phase coherence is lost because it is energetically favorable to suppress quantum fluctuations of number of particles in each site of an external periodic potential. Such a Mott insulator (MI) phase is characterized by a gap in the excitation spectrum \cite{fisher89,jaksch98}. Transition between the superfluid and Mott insulator phases has been demonstrated \cite{greiner02} and is extensively investigated in ultra-cold atom laboratories \cite{dutta15}. In the present letter it is shown that repulsive particle interactions of a periodically driven system lead to formation of Mott insulator-like state where long-time phase coherence is lost.

It has been proposed that spontaneous breaking of time translation symmetry can lead to formation of time crystals where probability density at a fixed point in the configuration space reveals spontaneously time periodic behaviour \cite{wilczek12,li12,chernodub12,wilczek13}. The possibility of such a spontaneous process is currently a subject of the debate in the literature \cite{bruno13,wilczek13a,bruno13a,li12a,bruno13b,nozieres13,watanabe14,sacha14}. In our study we do not consider the problem of the time crystal formation but concentrate on an analysis of systems where time-periodicity is already given by an external driving. 


\section*{Results}

\begin{figure}
\begin{center}
\resizebox{0.45\columnwidth}{!}{\includegraphics{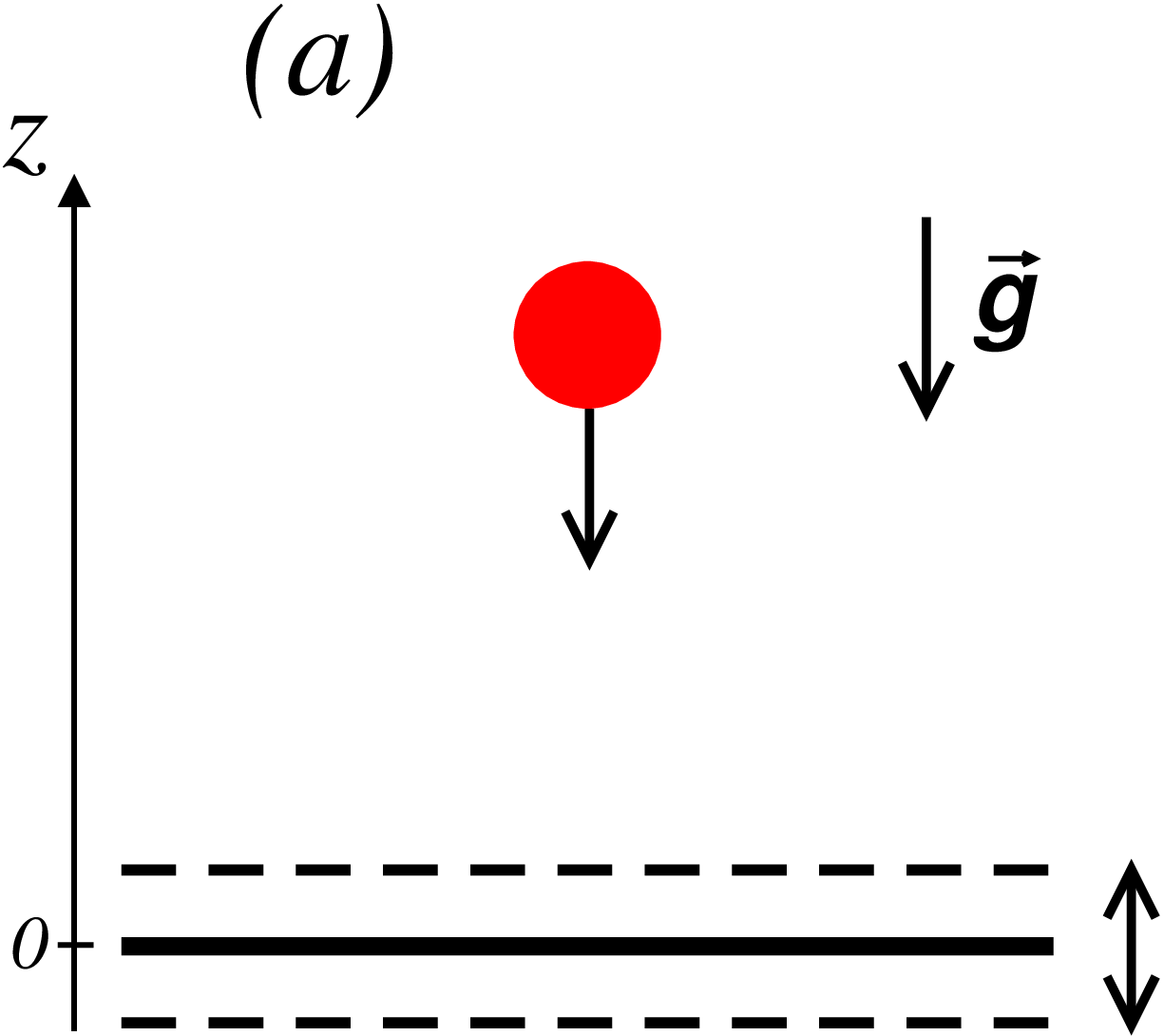}}
\resizebox{0.45\columnwidth}{!}{\includegraphics{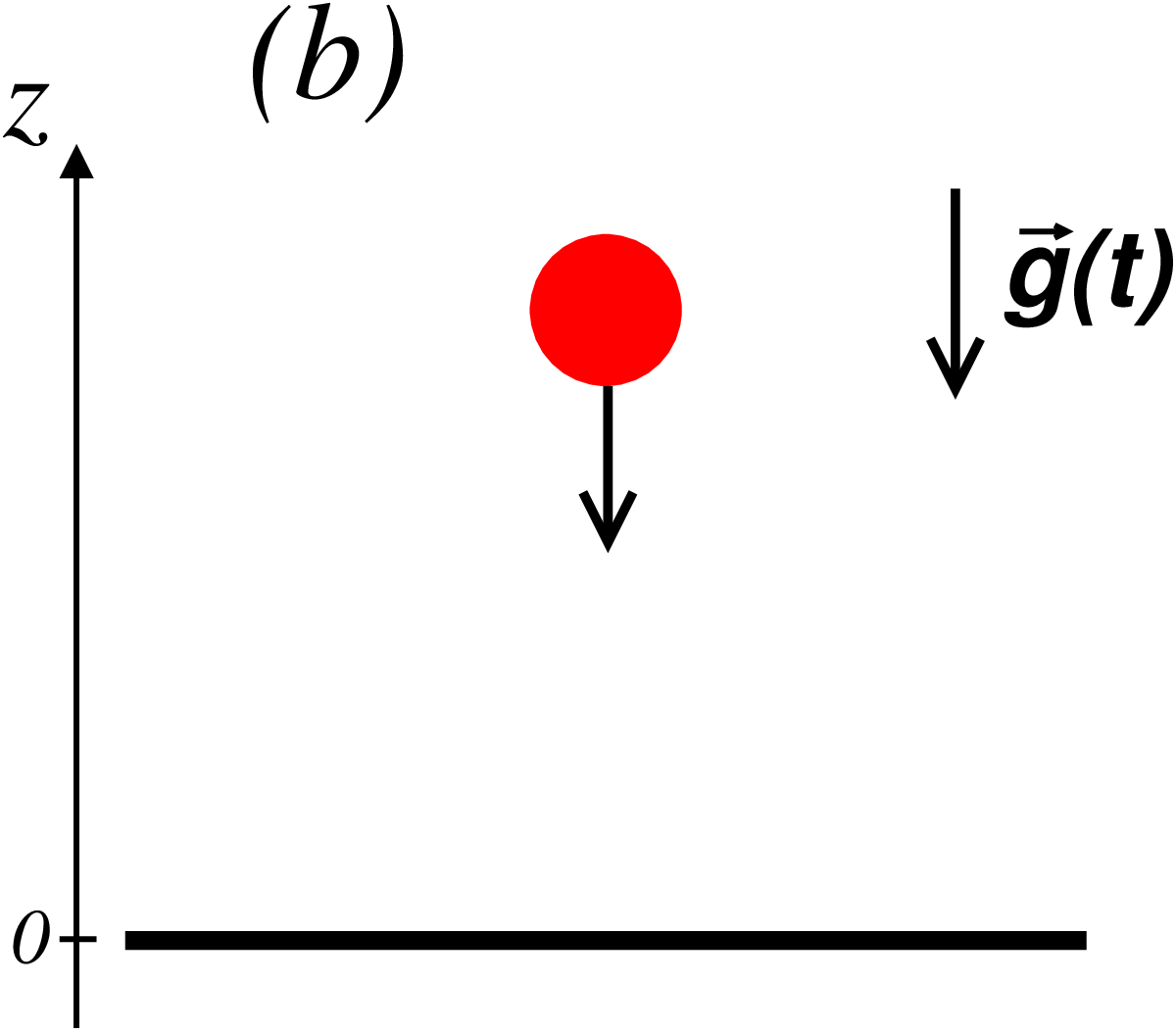}}
\caption{Panel (a): schematic plot of a particle bouncing on an oscillating mirror in the presence of the gravitational field. In the absence of the oscillations all classical trajectories are periodic with a period the longer, the greater energy of a particle. When the time periodic oscillations are on and their amplitude is not very big, orbits whose period is multiple of the mirror oscillation period survive. Such perturbation resistant orbits are located in resonant islands in the classical phase space. If the period of an orbit is $s$ times longer than the mirror oscillation period $T$, where $s$ is integer, we will call the orbit $s$-resonant. Panel~(b): description of the system is more convenient if we choose the coordinate frame that oscillates with the mirror \cite{buchleitner02}. Then, the mirror does not move but the gravitational acceleration becomes time dependent. The resulting Hamiltonian of the system reads $H(t)=p^2/2+z+\lambda z\cos(2\pi t/T)$ where the gravitational units have been used, i.e. $l_0=(\hbar^2/m^2g)^{1/3}$, $t_0=(\hbar/mg^2)^{1/3}$ and $E_0=mgl_0$ for length, time and energy, respectively, where $m$ is particle mass and $g$ original gravitational acceleration. The results presented in Figs.~\ref{fs}-\ref{al} correspond to $\lambda=0.01$ and $T=7.79$.
}
\label{class}
\end{center}
\end{figure}

\noindent
{\bf Anderson localization in the time domain.}
We consider single-particle systems that periodically depend on time, i.e. Hamiltonians $H(t+T)=H(t)$ where $T$ is the time period. While the energy is not conserved, there are so-called quasi-energy states $u_n$ (in analogy to Bloch states in spatially periodic problems) that are time-periodic eigenstates, $H_Fu_n=\varepsilon_nu_n$, of the so-called Floquet Hamiltonian $H_F(t)=H(t)-i\hbar\partial_t$. There are many different quasi-energy eigenstates of the systems but we will be interested in those that are represented by wave-packets localized on classical $s$-resonant orbits, i.e. on orbits whose periods equal $sT$ where $s$ is integer. Such extraordinary states exist in different experimentally attainable systems like, e.g., electronic motion in a hydrogen atom in microwave field \cite{delande94,birula94}, rotating molecules \cite{birula96} or an atom bouncing on an oscillating mirror \cite{buchleitner02}. The latter system will serve as an illustration of the ideas presented in this letter. In Fig.~\ref{class} classical motion of this system is described. From the semiclassical point of view, existence of wave-packets which move along classical orbits is related to localization of quantum states inside elliptical resonant islands in the phase space \cite{buchleitner02}. 

\begin{figure}
\begin{center}
\resizebox{0.9\columnwidth}{!}{\includegraphics{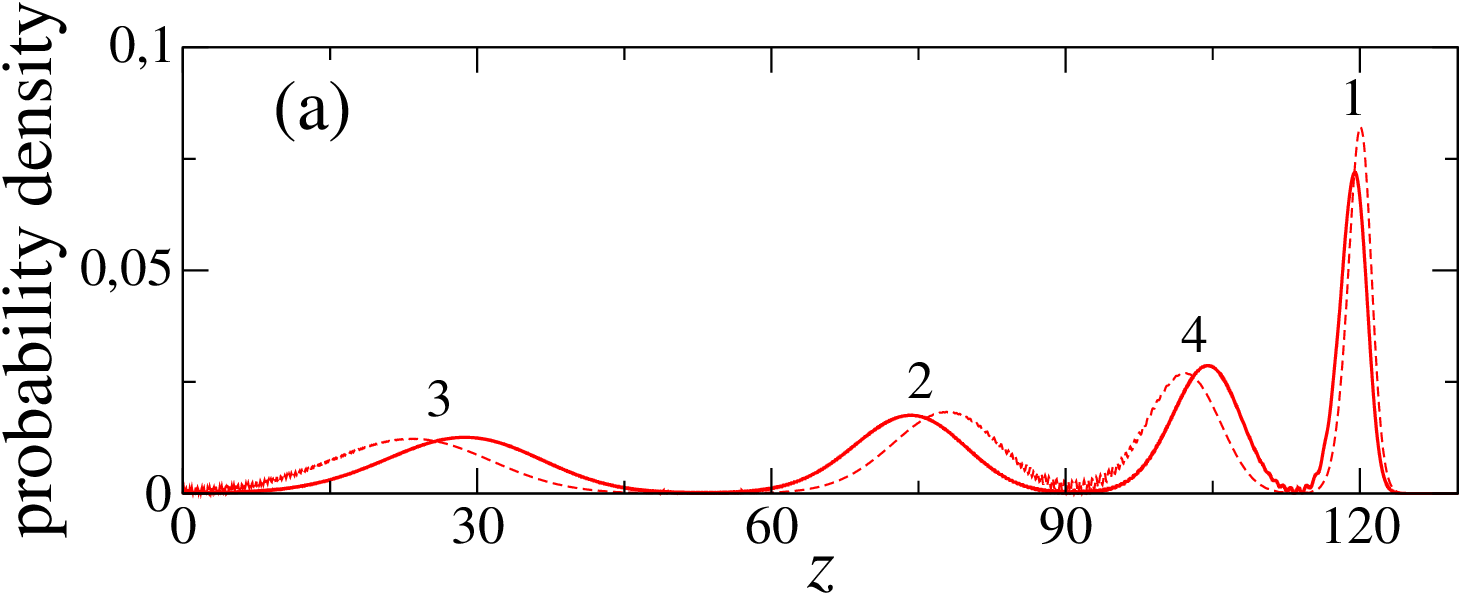}}
\resizebox{0.9\columnwidth}{!}{\includegraphics{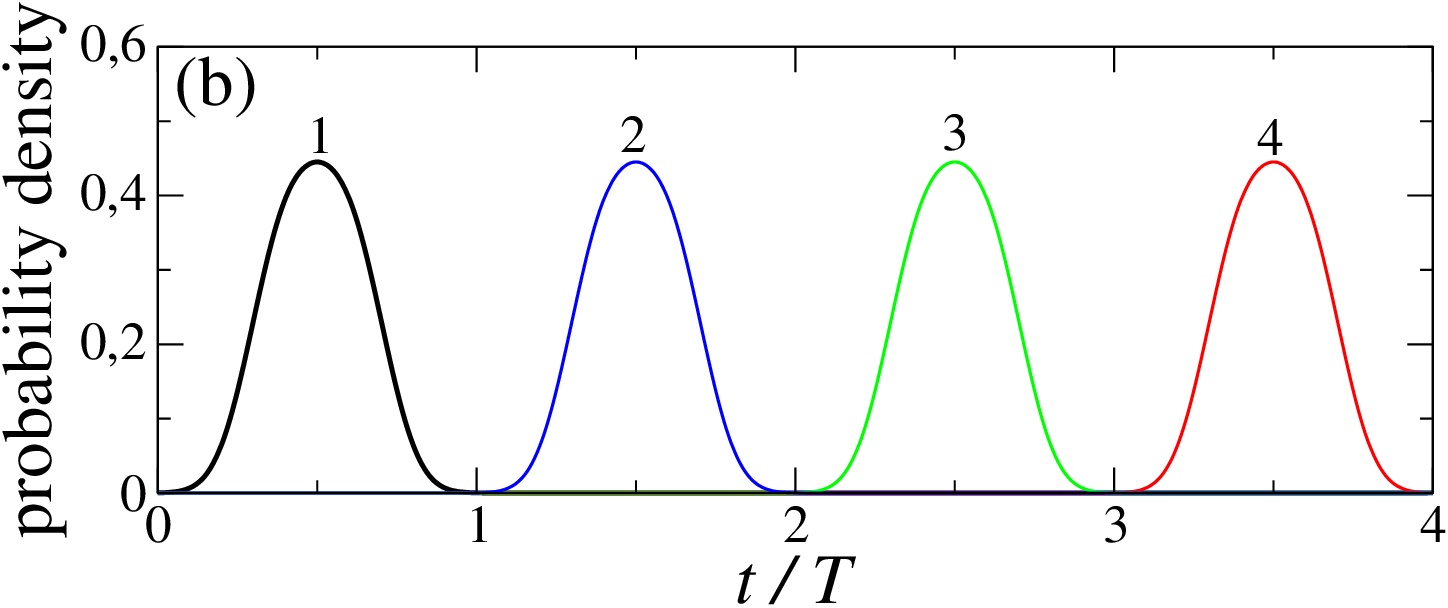}}
\caption{Analysis of the quasi-energy eigenstates localized on $s$-resonant periodic orbit, with $s=4$, for a particle bouncing on an oscillating mirror -- the classical behaviour is described in Fig.~\ref{class}. Panel (a) shows one of the eigenstates at $t=0.25T$ (solid line) and $t=0.3T$ (dash line). The eigenstate is a superposition of 4 wave-packets that move on the classical orbit. Each wave-packet evolves with the period $4T$ but after $T$ they exchange their positions what makes the whole eigenstate periodic with the period $T$. There are 4 such eigenstates that are localized on the orbit. Proper superpositions of the eigenstates allows one to extract 4 individual wave-packets, $\phi_j$, that are numbered in (a) and (b). After each period $T$ of the time evolution the wave-packets exchange their positions in the following order $\phi_{j+1}\rightarrow\phi_{j}$ (with $j+1$ modulo 4). Crystalline structures are not visible in the configuration space, however, they emerge in the time domain. Panel~(b) shows time evolution of the 4 wave-packets (plotted with different colours), whose superpositions form the eigenstates localized on the $4$-resonant orbit, at $z=121$ that is close to the turning point of a particle in the classical description. These wave-packets are analogues of Wannier states corresponding to the lowest energy band of a particle in a spatially periodic potential. Tunneling between wave-packets that are neighbours in the time domain are the leading tunneling processes. These processes are taken into account in (\ref{ef}).  The corresponding absolute values of the tunneling rates $|J_j|=2|\langle\langle\phi_{j+1}|H_F|\phi_j\rangle\rangle|=J\approx 6\times 10^{-7}$ for the system parameters described in Fig.~\ref{class}.
}
\label{fs}
\end{center}
\end{figure}

A single wave-packet moving on a $s$-resonant orbit cannot form a quasi-energy eigenstate because its period of motion is $s$ times longer than the required period for all eigenstates of the Floquet Hamiltonian. However, superpositions of $s$ wave-packets can form system eigenstates -- in Fig.~\ref{fs} we show an example for the $s=4$ case. There are $s$ such superpositions that are linearly independent, hence, there are $s$ eigenstates which reveal localized wave-packets moving along a $s$-resonant orbit. The corresponding quasi-energy levels $\varepsilon_n$ are nearly degenerated with only small splittings related to rates of tunneling of individual wave-packets. That is, if a single wave-packet is prepared initially on a $s$-resonant orbit, it travels along the orbit but after some time tunnels to the positions of other wave-packets which form the system eigenstates but which are missing initially. If $s\rightarrow\infty$, the nearly degenerated eigenvalues $\varepsilon_n$ form a quasi-energy band which is an analogue of the lowest energy band of spatially periodic systems. Now we can establish conjecture about the behaviour of periodically driven systems and systems of particles in spatially periodic potentials. A single wave-packet localized on a $s$-resonant orbit is an analogue of a Wannier state localized in a single site of a spatially periodic potential \cite{dutta15} but in the time domain, see Fig.\ref{fs}. There exist also excited wave-packets that move on a $s$-resonant trajectory \cite{buchleitner02} which are analogues of Wannier states corresponding to excited energy bands of spatially periodic systems. The conjecture becomes formal when we derive Floquet energy of a periodically driven particle in the Hilbert subspace spanned by $s$ individual wave-packets $\phi_j$,
\bea
E_F&=&\langle\langle\psi|H_F|\psi\rangle\rangle
\approx -\frac12\sum\limits_{j=1}^{s}(J_{j}a_{j+1}^*a_j+{\rm c.c.}),
\label{ef}
\eea
where  $\langle\langle\psi|H_F|\psi\rangle\rangle=\int\limits_0^{sT}dt\langle\psi|H_F(t)|\psi\rangle$ and $\psi\approx\sum\limits_{j=1}^sa_j\phi_j$ has been substituted \cite{sacha14}. The tunneling rates between wave-packets that are neighbours in the time domain (see Fig.~\ref{fs}) $J_j=-2\langle\langle\phi_{j+1}|H_F|\phi_j\rangle\rangle$ with the same absolute value $|J_j|=J$.

\begin{figure}
\begin{center}
\resizebox{0.9\columnwidth}{!}{\includegraphics{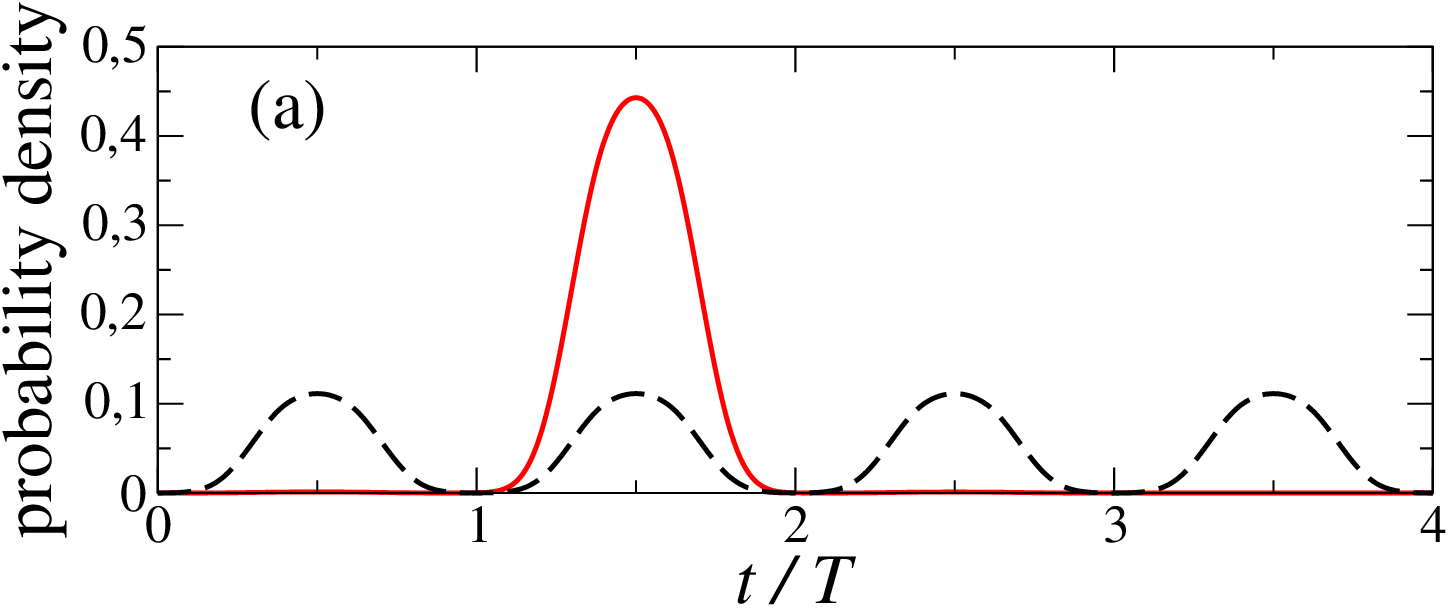}}
\resizebox{0.9\columnwidth}{!}{\includegraphics{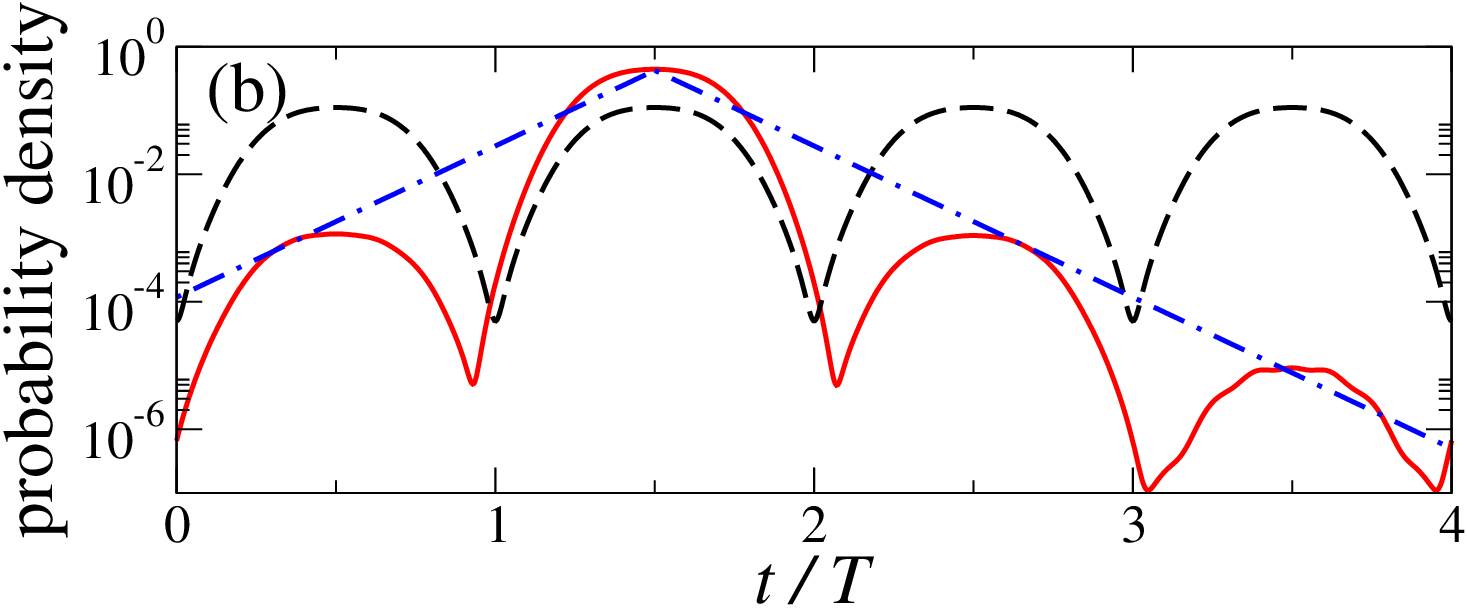}}
\caption{Quantum description of a particle moving on the 4-resonant orbit in the presence of a small perturbation, $H'(t)=z\sum_{n=1}^4\alpha_n\cos\left\{2\pi\left[n\frac{t}{4T}-\sin\left(2\pi \frac{t}{4T}\right)\right]\right\}$, that fluctuates in time. The whole Floquet Hamiltonian $H_F(t)+H'(t)$ is time periodic with the period $4T$ and so do the eigenstates. The coefficients $\alpha_n$, in $H'$, are chosen so that the set of  $\langle\langle\phi_j|H'|\phi_j\rangle\rangle$ reproduces a chosen set of numbers $E_j$, where $\phi_j$'s are the wave-packets described in Fig.~\ref{fs}. We have chosen $E_j=4Jw_j$ where $w_j$ are random numbers corresponding to a Lorentzian distribution $W(w_j)=1/(\pi+\pi w_j^2)$. Then, the Floquet energy (\ref{ef}) supplemented with $E'=\sum_j E_j|a_j|^2$ constitutes the Lloyd model of 1D lattice where all eigenstates are Anderson localized and the exact expression for the localization length is known \cite{haake2010}. From the Lloyd model we know that the eigenstates are superpositions of the wave-packets, $\sum_ja_j\phi_j(z,t)$, with $|a_j|^2\propto e^{-|j_0-j|/l}$ where $l$ is the AL length and $j_0$ is a number of the wave-packet around which a given eigenstate is localized. The wave-packets $\phi_j$ arrive at a given position $z$ in equidistant intervals in time, thus, the AL length in time is $l_t=lT$. Solid lines in (a) and (b) show one of the 4 eigenstates localized on the 4-resonant orbit at $z=121$ versus time in the linear (a) and logarithmic (b) scales (the eigenstate is obtained in full numerical diagonalization of $H_F+H'$ but the solution of the Lloyd model is identical). Dash lines present behaviour of the eigenstates in the absence of the disorder in time, i.e. when $H'=0$. 
Despite the fact that the system is rather small, the characteristic exponential decay of the humps is clearly visible in (b) -- the fitted exponential profile (dash-dotted line) corresponds to $l_t=0.18T$.
}
\label{al}
\end{center}
\end{figure}

Description of a resonantly driven particle has been reduced to the problem of a particle in 1D lattice with nearest neighbour tunnelings (\ref{ef}). If we are able to create an additional disorder term, $E'=\sum\limits_{j=1}^sE_j|a_j|^2$ where $E_j$'s are random numbers, Eq.~(\ref{ef}) will become the 1D Anderson model and AL phenomena will emerge. The disorder term can be realized by an additional small perturbation $H'(t)$ that fluctuates in time. That is, $H'(t+sT)=H'(t)$ but between $t$ and $t+sT$ it behaves so that the set of $E_j=\langle\langle\phi_j|H'|\phi_j\rangle\rangle$ reproduces a chosen set of random numbers. Then, the quasi-energy eigenstates of the total system, $H_F(t)+H'(t)$, are time periodic with the period $sT$ but reveal superpositions of individual wave-packets $\phi_j$ with exponentially localized distributions. In Fig.~\ref{al} we describe an example of the realization of time disorder and show solutions that exhibit AL. 

Time crystal phenomena are related to time-periodic evolution of probability density for  measurement of a particle at a fixed position. If there is no time disorder, such a probability reveals a uniform train of $s$ humps that is repeated every $sT$ period, see Fig.~\ref{fs}(b). If the disorder is on, the train is no longer uniform but reveals an Anderson localized distribution of humps, see Fig.~\ref{al}.


\begin{figure}
\begin{center}
\resizebox{0.43\columnwidth}{!}{\includegraphics{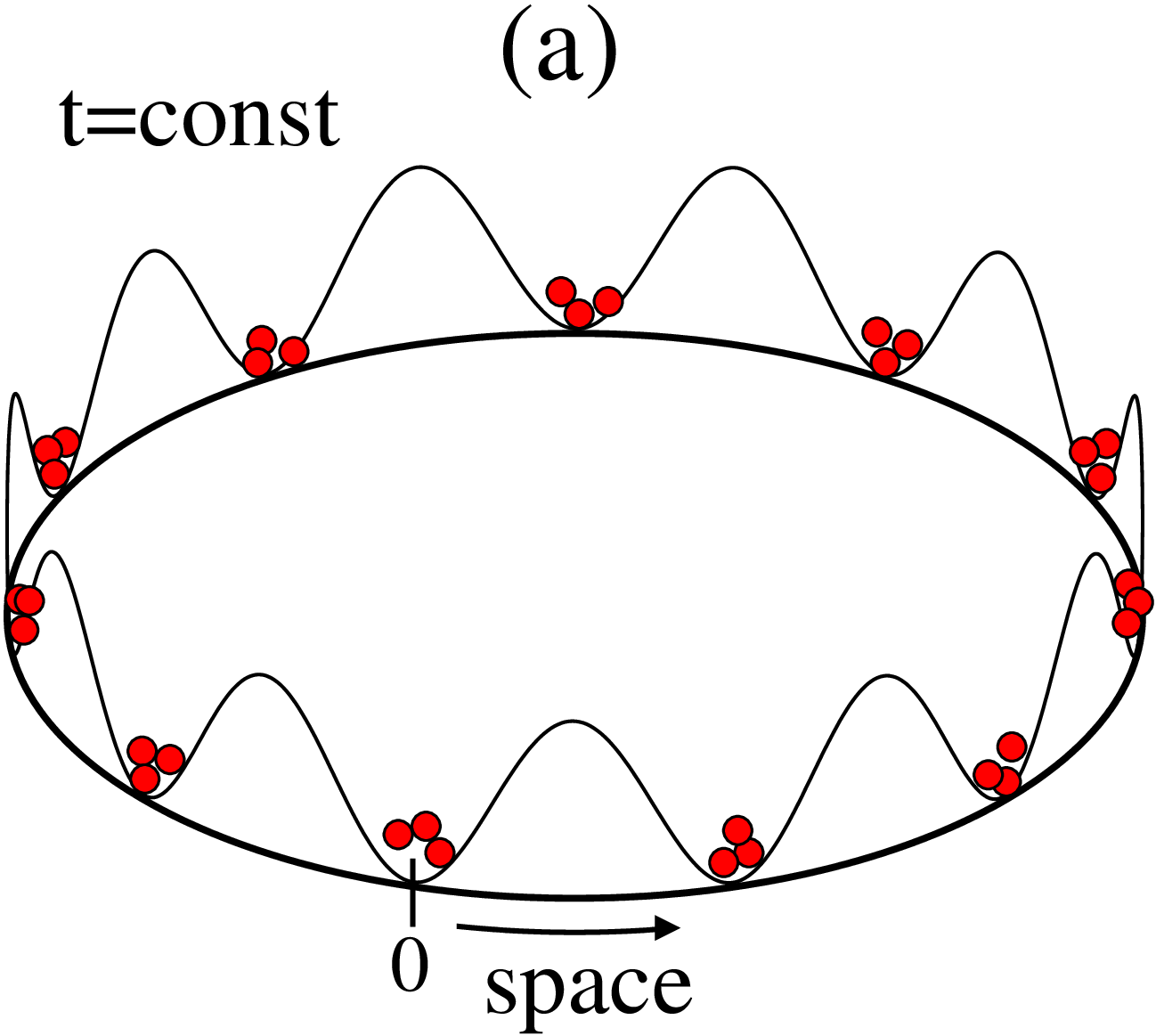}} 
\hfill
\resizebox{0.43\columnwidth}{!}{\includegraphics{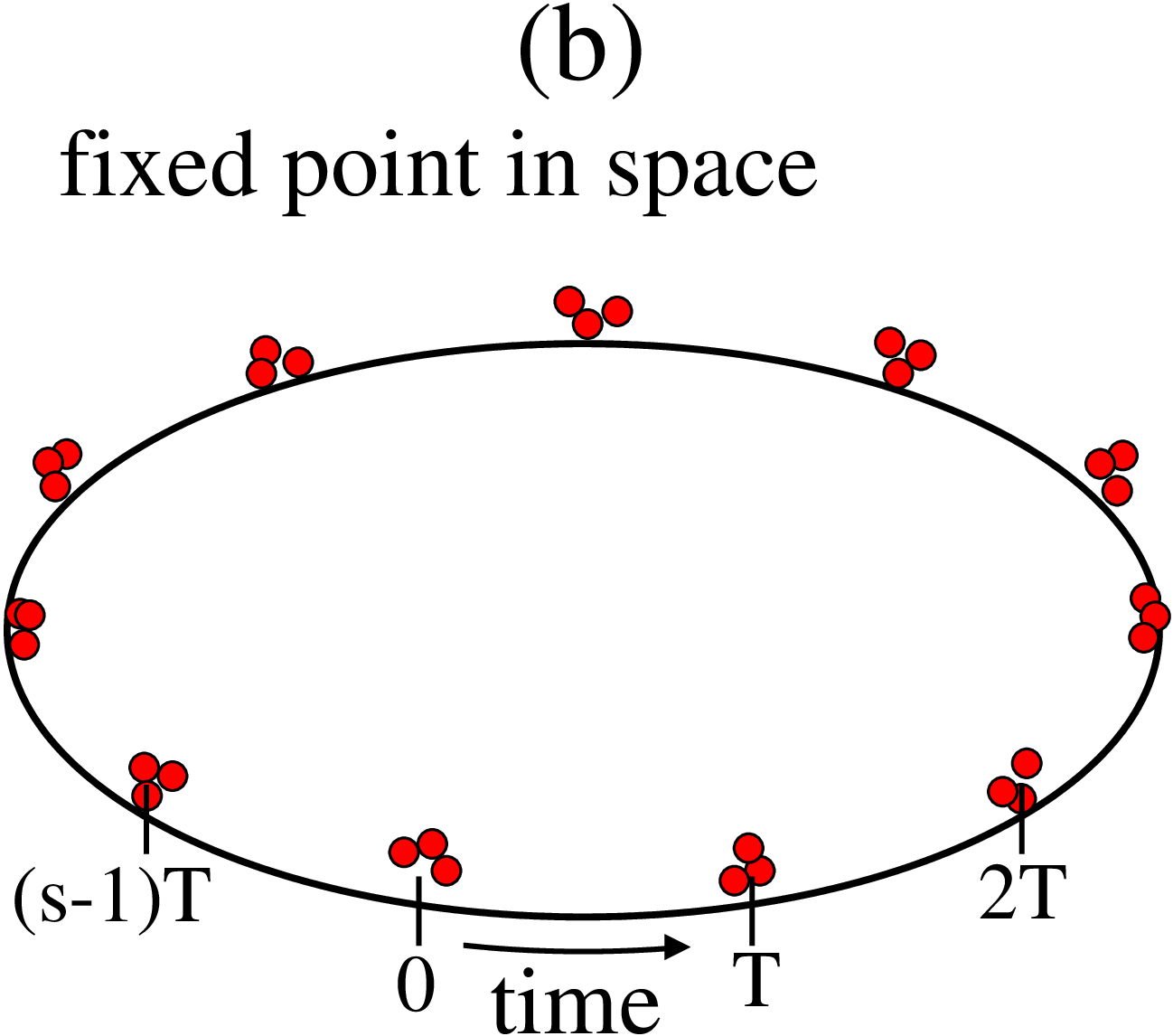}}
\caption{Comparison of crystalline structures in the space and time domains. Panel (a) illustrates a space periodic potential in 1D with the periodic boundary conditions, the larger the system, the greater number of potential sites along the ring. Particles in the potential wells illustrate a Mott insulator state where in each potential site there is a well defined number of particles. Panel (b) refers to a crystal structure in the time domain with the time periodic conditions corresponding to the period $sT$. In the Mott insulator regime, at a fixed position, a well defined numbers of particles arrive every time interval $T$ like on a conveyor belt or in a machine gun. The greater $s$, the larger time crystal is.
}
\label{mott}
\end{center}
\end{figure}

\noindent
{\bf Mott insulator phase in the time domain.}
Let us consider $N$ bosonic particles with repulsive contact interactions, characterized by a parameter $g_0>0$, that are periodically driven, e.g. we can focus on ultra-cold atoms bouncing on an oscillating mirror. In the absence of the interactions, a Bose-Einstein condensate can be formed. Then, the system description reduces to a single particle problem and $s$-resonant driving can be described by the Floquet energy (\ref{ef}). In order to describe behaviour of the interacting many-body system we may truncate the Hilbert space to a subspace spanned by Fock states $|n_1,\dots,n_s\rangle$ where the occupied modes correspond to localized wave-packets $\phi_j$ moving along a $s$-resonant trajectory. Then, the many-body Floquet Hamiltonian reads 
\be
\hat {\cal H}_F\approx -\frac12\sum\limits_{i=1}^{s}(J_{i}\hat a_{i+1}^\dagger \hat a_i+{\rm h.c.})+\frac12\sum\limits_{i,j=1}^{s}U_{ij}\hat n_i\hat n_j,
\label{bh}
\ee
where $\hat a_i$ and $\hat a_i^\dagger$ are bosonic anihilation and creation operators and $\hat n_i=\hat a_i^\dagger\hat a_i$. The coefficients $U_{ij}=g_0\langle\langle\phi_i|\phi_i\phi_j^*|\phi_j\rangle\rangle$ describe interactions between particles that ocupy the same mode (for $i=j$) and between particles in different modes ($i\ne j$). The latter are negligible if propagating wave-packets never overlap. If they pass each other during the evolution along the orbit, $U_{ij}\ne 0$ and the Hamiltonian (\ref{bh}) corresponds to a lattice model with long range interactions what could be surprising because the original particle interactions are zero range. For the parameters described in Fig.~\ref{class} and $s=4$, $U_{ij}$ are about $0.1U_{ii}$ for $i\ne j$. 

If $g_0\rightarrow 0$, the ground state of (\ref{bh}) is a superfluid state with long-time phase coherence. However, the long-time phase coherence is lost when $U_{ii}\gg NJ/s$ because the ground state becomes a Fock state $|N/s,N/s,\dots,N/s\rangle$ \cite{jaksch98,dutta15}. If we fix position in the configuration space and analyze what is the phase relation between wave-packets arriving one by one at this position (see Fig.~\ref{mott}), it turns out the phase is totally undefined. Moreover, there is a gap in the excitation spectrum of the order of $U_{ii}$. Hence, the system reveals Mott insulator-like properties but in the time domain. Ultra-cold dilute atomic gases are promising laboratories for the realization of such a MI phase due to an unprecedented level of experimental 
control.

\section*{Discussion}

\noindent
Summarizing, we have shown that periodically driven systems can reveal non-trivial crystalline properties in the time domain. One-particle systems show Anderson localization in the time evolution if there are fluctuations in the periodic driving. Long-time phase coherence of periodically driven many-body systems can be lost due to particle interactions and the Mott insulator-like phase emerges.

\section*{Acknowledgments}

\noindent
Support of Polish National Science Centre via project number DEC-2012/04/A/ST2/00088 is acknowledged. The work was performed within the project of Polish-French bilateral programme POLONIUM and the FOCUS action of Faculty of Physics, Astronomy and Applied Computer Science of Jagiellonian University. 


%

\end{document}